\documentclass[runningheads]{llncs}
\usepackage{graphicx}
\usepackage{verbatim}
\usepackage{placeins}

\begin{document}
\title{Estimation of Cardiac Valve Annuli Motion with Deep Learning}
%
%

\author{
Eric Kerfoot \inst{1} \and
Carlos Escudero King \inst{1} \and
Tefvik Ismail \inst{1} \and
David Nordsletten \inst{1,2} \and
Renee Miller \inst{1}}

\authorrunning{E. Kerfoot et al.}
%

\institute{King's College London, London, UK \and
University of Michigan, Ann Arbor, USA \\
\email{\{eric.kerfoot, carlos.escudero, tefvik.ismail, david.nordsletten, renee.miller\}@kcl.co.uk}\\
}

\maketitle              
\begin{abstract}

Valve annuli motion and morphology, measured from non-invasive imaging, can be used to gain a better understanding of healthy and pathological heart function. Measurements such as long-axis strain as well as peak strain rates provide markers of systolic function. Likewise, early and late-diastolic filling velocities are used as indicators of diastolic function. Quantifying global strains, however, requires a fast and precise method of tracking long-axis motion throughout the cardiac cycle. Valve landmarks such as the insertion of leaflets into the myocardial wall provide features that can be tracked to measure global long-axis motion. Feature tracking methods require initialisation, which can be time-consuming in studies with large cohorts. Therefore, this study developed and trained a neural network to identify ten features from unlabeled long-axis MR images: six mitral valve points from three long-axis views, two aortic valve points and two tricuspid valve points. This study used manual annotations of valve landmarks in standard 2-, 3- and 4-chamber long-axis images collected in clinical scans to train the network. The accuracy in the identification of these ten features, in pixel distance, was compared with the accuracy of two commonly used feature tracking methods as well as the inter-observer variability of manual annotations. Clinical measures, such as valve landmark strain and motion between end-diastole and end-systole, are also presented to illustrate the utility and robustness of the method.

\keywords{Cardiac valve motion \and Landmark detection \and Regression network \and Deep learning}
\end{abstract}
\section{Introduction}


Cardiovascular disease in patients can result in reduced quality of life and premature mortality, becoming increasingly prevalent with age \cite{RN2}. The integrity of all four cardiac valves, mitral, tricuspid, aortic and pulmonary, is crucial for appropriate haemodynamic function. 
The annulus is a ring-shaped structure that forms the perimeter of the valve opening, and its shape changes dynamically during the cardiac cycle \cite{RN5}. Characterising the physiological deformation of the valve annuli and how their shapes change in cardiac pathologies has become a topic of increased research. For example, excessive dilation of the aortic valve annulus has been identified as a predictor of recurrent aortic regurgitation following surgical intervention in aortic valve disease \cite{RN7}. Further, the mitral valve annulus has been shown to be ‘saddle-shaped’ in healthy individuals, but becomes increasingly dilated and flattened during systole in mitral regurgitation \cite{RN26}. Many previous studies have focused on the role of a single valve annulus in cardiac function. However, simultaneous characterisation of the motion and morphology of all four valves could provide a more comprehensive view of their interactions throughout the cardiac cycle.
Long-axis motion and strain is used as an indicator of systolic function. For example, mitral/tricuspid annular plane systolic excursion (MAPSE/ TAPSE), measured as the distance that the lateral valve landmark moves between end-diastole and end-systole, can provide a quantitative marker of systolic function, particularly in patients who have impaired left ventricular function despite a normal ejection fraction (e.g. patients with hypertrophic cardiomyopathy, HCM) \cite{Bulluck2014}. Long-axis strain, quantified using annotations of valve landmarks, provides a normalized metric for assessing cardiac function \cite{Brecker2000}. One recent study showed that long-axis strain measurements obtained by a semi-automated method of tracking valve positions was able to differentiate patients with heart failure from controls \cite{Leng2020}. 


Previous approaches to measuring the motion of valve annuli from cardiac magnetic resonance (MR) have utilised semi-automated feature tracking methods \cite{RN5}. In principle, most cardiac MR tracking tools follow the features of the blood-myocardium boundary using a form of a maximum likelihood estimation \cite{Schuster2016}. Tracking methods require manual initialisation of the feature or boundary to be tracked and have been shown to have low reproducibility \cite{Schuster2015}. However, methods exist which integrate machine learning methods, such as marginal space learning and trajectory spectrum learning, to accurately capture the shape and deformation of valves through the cardiac cycle \cite{Ionasec2010}. 

The aim of this study is to train a single deep learning neural network to robustly identify 10 valve landmarks, describing the positions of three out of the four valves, from three different long-axis MR image orientations which are acquired as part of a standard clinical routine. 
The valve landmarks, identified through the entire cardiac cycle, are then used to characterise motion in a diverse cohort of patients and healthy controls. We anticipate that this tool can provide a method for rapidly assessing valve annulus motion and morphology, which does not require manual initialisation or input. Ultimately, applied to a large cohort, this tool can provide further understanding of the interaction and role of the valve annuli in healthy and pathological cardiac function. 

\section{Methods}

\subsection{Training Data and Manual Annotation}

\begin{figure} 
\centering
\includegraphics[width=\textwidth]{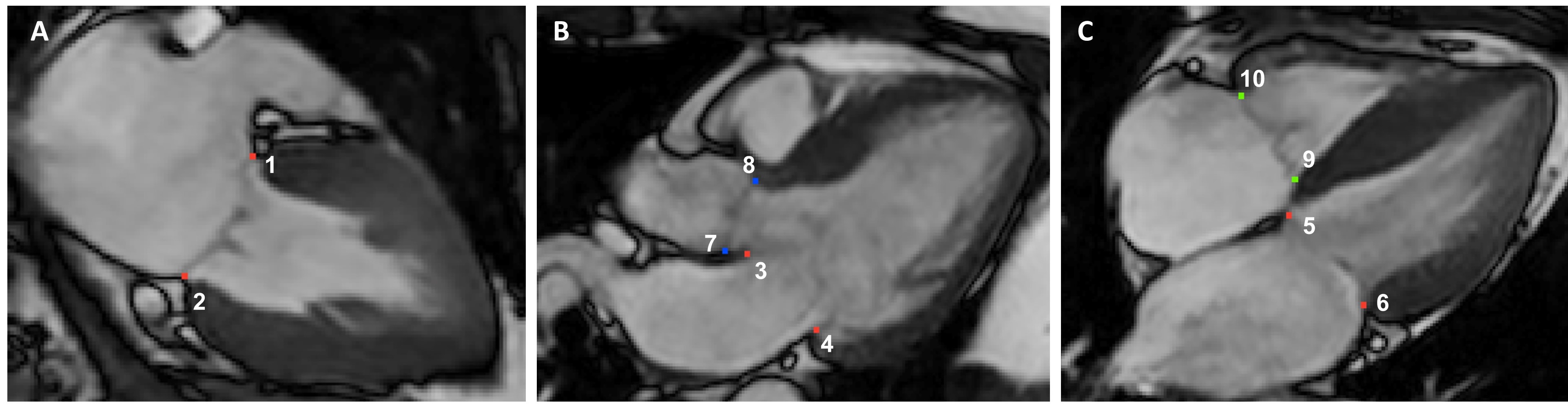}
\caption{Manually annotated valve landmarks in the A) 2-chamber, B) 3-chamber, and C) 4-chamber long-axis images labelled from 1-10. Red: mitral valve, blue: aortic valve, green: tricuspid valve.} 
\label{fig:valve-labels}
\end{figure}

The training dataset consisted of 8574 long-axis images collected from hypertrophic cardiomyopathy (HCM, n = 3069) and myocardial infarction (MI, n = 5505) patients. 
All HCM patients were part of an ongoing imaging study. 
MI patients were a subset from the DETERMINE study obtained through the Cardiac Atlas Project (http://stacom.cardiacatlas.org/) \cite{RN27}. 
All images were acquired in the 2-chamber (2CH, n = 2952), 3-chamber (3CH, n = 2712) or 4-chamber (4CH, n = 2910) view on either Philips (n = 4463), Siemens (n = 3195) or GE scanners (n = 916). The validation set consisted of 930 images.
Variations in the positioning of these views occurred due to natural and pathological variations in patient heart shape and valve positioning. 
Generally, the 4CH image (Figure \ref{fig:valve-labels}C) shows the mitral and tricuspid valves, which control flow of blood into the left and right ventricles, respectively. 
The 3CH image (Figure \ref{fig:valve-labels}B) provides views of the mitral and aortic valves. 
Finally, the 2CH image (Figure \ref{fig:valve-labels}A) contains a view solely of the mitral valve. 
Each long-axis view was composed of approximately 30 frames through the entire cardiac cycle, with slight variation based on patient heart rate and breath-hold duration required during acquisition. 
Ten valve landmarks, including the mitral (1-6), aortic (7-8) and tricuspid (9-10) valves, were manually annotated at every frame by an expert reviewer to provide a ground-truth for training (Figure \ref{fig:valve-labels}). 
Image annotation was conducted using ITK-SNAP \cite{RN24}. Ten cases (960 images) were annotated by two additional expert observers to calculate the inter-observer variability.

\subsection{Feature Tracking}

As a comparison with the neural network prediction, valve landmarks were tracked using two common feature tracking algorithms available in Matlab 2019b (Kanade-Lucas-Tomasi algorithm) \cite{Tomasi1991} and MIRTK \cite{Schuh2014}. 

\subsection{Neural Network}

\begin{figure}[t]
\begin{center}
\includegraphics[width=\textwidth]{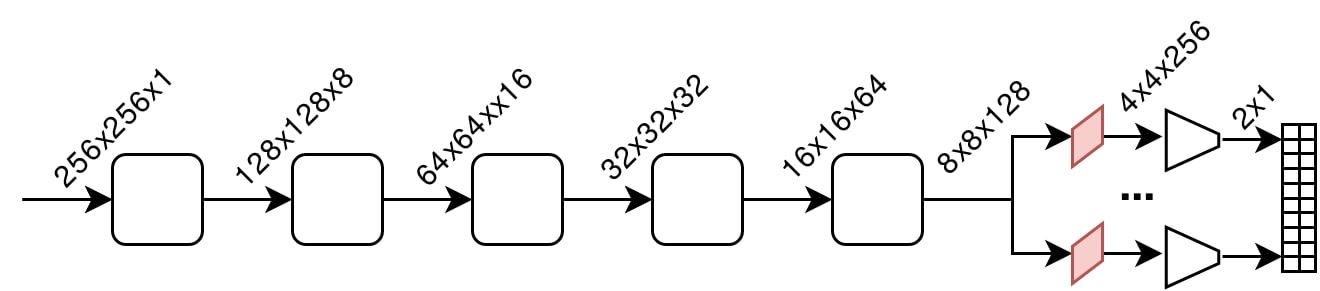}
\vspace{1mm}
\includegraphics[width=\textwidth]{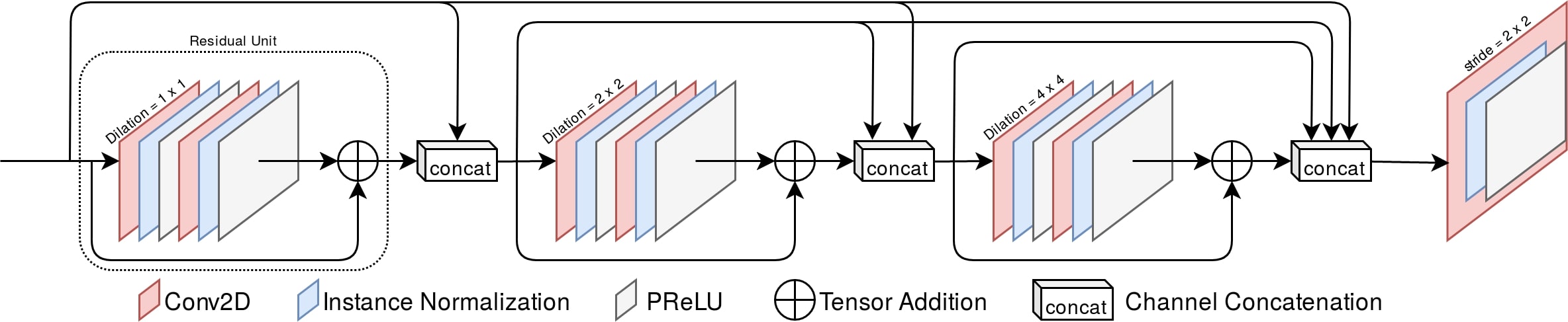}
\caption{
(Top) Regressor network topology. (Bottom)
Dilated dense block of convolutions used to construct the regressor network. 
This dense block is composed of three series of residual units whose convolutions have a dilation of 1, 2 and 4, respectively. 
The outputs from each unit is concatenated with that from the previous, such that the final volume is composed of features derived from viewing areas of different size. 
A final convolution is used to down-sample the output volume by a factor of two. 
All convolution kernels have 3x3 shape.} 
\label{fig:dilateddenseblock}
\end{center}
\end{figure}

The network architecture, shown in Figure \ref{fig:dilateddenseblock}, is composed of a sequence of dilated dense blocks~\cite{DBLP:journals/corr/HuangLW16a} represented by each square with input and output volume sizes given.
Each dense block applies 2D convolutions with different dilation rates to their inputs and concatenates the results from each together as its output.
A dense block produces an output volume containing feature information derived from multiple sizes of convolutional fields-of-view, subsequent convolutions can thus combine information at different scales to recognise small and large features.

The dimensions of the output volume from each dense block is half as large as the input in the spatial dimensions.
After five such blocks, the volume is passed to a series of small networks, each composed of a convolution followed by a fully-connected layer, which compute one landmark coordinate each from the volume.
These small networks can learn a specific reduction for each landmark, and are able to correctly learn to produce a result of $(0,0)$ for landmarks not present in the image.
The network was trained for 120,000 iterations using a straight-forward supervised approach with 8,574 image-landmark pairs.
The Adam optimizer~\cite{Kingma2014} was used with a learning rate of 0.0001, and the loss computed as the mean absolute error between ground truth and predicted landmarks.

Minibatches were collected together by choosing random images from the training set and applying a randomly-selected series of flip, rotation, shift and transpose operations to the image and landmark pairs. 
The images were further augmented by adding stochastic noise, applying a dropout function in k-space to partially corrupt the image, and varying the image intensity with a smooth shift function.
Lastly, free-form deformation was applied to the images and the landmarks to further vary the data. 
The result of these augmentations was to minimize over-fitting to the training data set and vastly increase the variation in features the network observed over the course of training.

During training, after every epoch of 400 iterations, a per-landmark accuracy value was calculated by applying the network to a validation data set of 930 image-landmark pairs.
For each landmark, the mean Cartesian distance between the set of ground truth and predicted coordinates was computed as the per-landmark prediction error. 
The minibatch selection procedure was then adjusted such that each image's probability of being selected was computed to be proportional to the distance error of the landmarks defined for it.
This step ensured that landmarks exhibiting lower accuracy would, at later stages in the training, become more common in the selected minibatches.
This curriculum learning approach~\cite{curriculum2009,10.1007/978-3-319-95921-4_26} forced the network to concentrate on those landmarks which were more difficult to accurately place compared to others, enforcing their importance to the overall loss calculation.

The trained neural network was then used to predict valve landmarks in long-axis images from patients with HCM, MI, and dilated cardiomyopathy (DCM), as well as healthy volunteers (Vol). 
Long-axis strain was calculated as the normalised distance from a stationary apical point to the valve landmarks. 
Mitral/tricuspid annular plane systolic excursion (MAPSE/TAPSE) values, calculated as the distance between lateral valve landmarks (6 and 10) in the 4CH view at end-diastole and end-systole, were also computed for each group.

\section{Results and Discussion}

\begin{figure}[t]
\includegraphics[width=\textwidth]{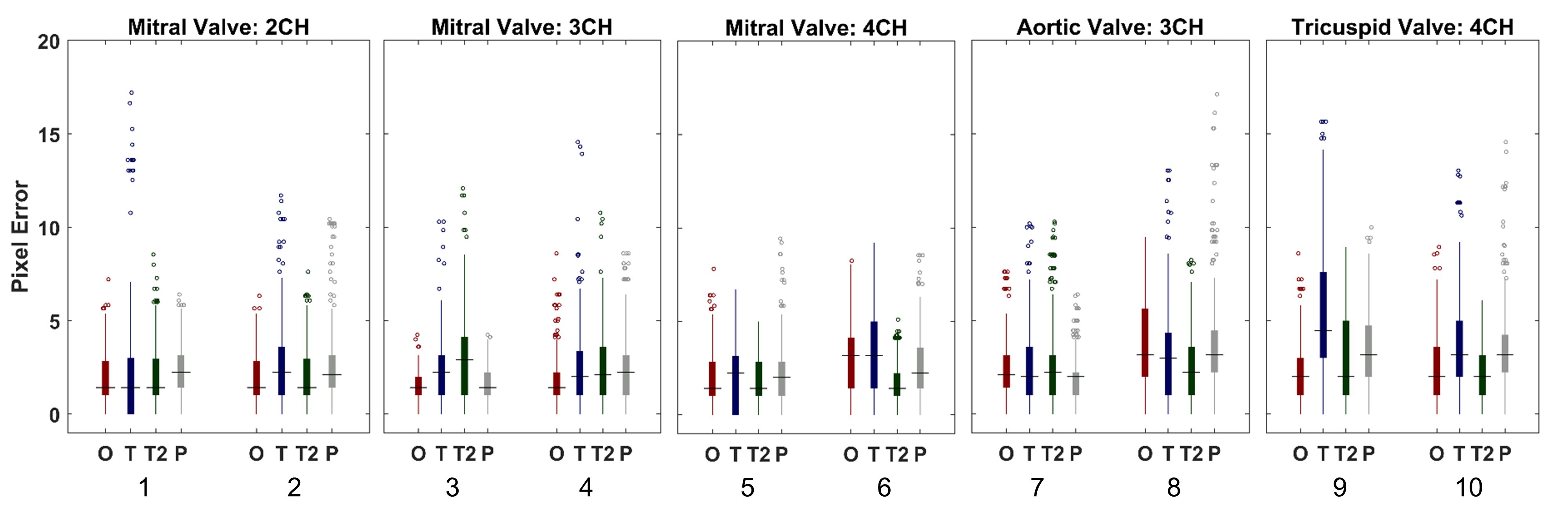}
\caption{Boxplots show the differences between inter-observer (O - red), Kanade-Lucas-Tomasi tracking (T - blue), MIRTK tracking (T2 - green) and neural network predicted landmarks (P - grey) errors for all valve landmarks (1-10). Lines represent the median, boxes the interquartile ranges and circles are outliers.} 
\label{fig:all-error}
\end{figure}

Boxplots in Figure \ref{fig:all-error} illustrate the differences between inter-observer, tracking and predicted point errors for each valve landmark. All mean pixel errors ($\pm$ one standard deviation) are shown in Table \ref{table:mean-error}. Mean inter-observer error was lower than mean predicted point errors in all landmarks except 7 and 8 (aortic valve). 

\begin{table}
\centering
\caption{Mean ($\pm$ one standard deviation) pixel error for the interobserver error (O), Kanade-Lucas-Tomasi (KLT) tracking (T), MIRTK tracking (T2) and neural network predicted landmarks (P).}\label{tab1}
\begin{tabular}{|c|c|c|c|c|}
\hline
Landmark & Inter-Observer &  KLT Tracking & MIRTK Tracking & Neural network \\
\hline
1 & 2.013 $\pm$ 1.571 & 2.311 $\pm$ 3.017 & 1.979 $\pm$ 1.764 & 2.334 $\pm$ 1.398 \\
2 & 2.422 $\pm$ 1.791 & 2.640 $\pm$ 2.307 & 1.927 $\pm$ 1.650 & 2.560 $\pm$ 2.154 \\
3 & 1.463 $\pm$ 1.089 & 2.375 $\pm$ 1.745 & 2.861 $\pm$ 2.297 & 1.773 $\pm$ 0.905 \\
4 & 2.024 $\pm$ 1.070 & 2.458 $\pm$ 2.167 & 2.367 $\pm$ 1.945 & 2.401 $\pm$ 1.704 \\
5 & 1.684 $\pm$ 1.470 & 2.169 $\pm$ 1.915 & 1.851 $\pm$ 1.193 & 2.270 $\pm$ 1.571 \\
6 & 2.717 $\pm$ 1.808 & 3.204 $\pm$ 2.210 & 1.619 $\pm$ 1.251 & 2.758 $\pm$ 1.763 \\
7 & 3.042 $\pm$ 2.948 & 2.309 $\pm$ 2.149 & 2.683 $\pm$ 2.357 & 1.993 $\pm$ 1.172 \\
8 & 4.311 $\pm$ 3.086 & 3.051 $\pm$ 2.692 & 2.422 $\pm$ 1.789 & 3.765 $\pm$ 2.691 \\
9 & 1.949 $\pm$ 1.505 & 5.411 $\pm$ 3.661 & 2.666 $\pm$ 2.371 & 3.545 $\pm$ 2.213 \\
10 & 3.440 $\pm$ 2.832 & 3.724 $\pm$ 2.636 & 2.140 $\pm$ 1.632 & 3.522 $\pm$ 2.235 \\
\hline
\end{tabular} \label{table:mean-error}
\end{table}

Although mean errors for the network-predicted landmarks were higher than inter-observer errors, the mean errors were less than or comparable to errors from the tracking methods. Tracking methods require manual initialisation whereas the neural network prediction requires no user input. Additionally, the network is able to not only provide accurate point locations but also accurate labels (e.g. 1-10), meaning that not only the long-axis view but also the image orientation is correctly identified by the neural network. For these reasons, the neural network could be used as a stand-alone tool or in conjunction with a more robust tracking method as a tool to automatically initialise landmark locations. 

The current network predicts the presence and location of 10 valve landmarks in individual long-axis images, not incorporating any temporal consistency. Therefore, time-series images are not needed and valve landmarks can be obtained from a single frame. However, incorporating temporal consistency may prove beneficial in cases with poor image quality in individual frames, for example, due to flow artefacts during the ejection phase. For the moment, post-processing steps can be used to interpolate landmark positions in these frames. The addition of temporal consistency will be evaluated in future work. 

\subsection{Clinical Measurements}

\begin{figure}[t]
\centering
\includegraphics[width=\textwidth]{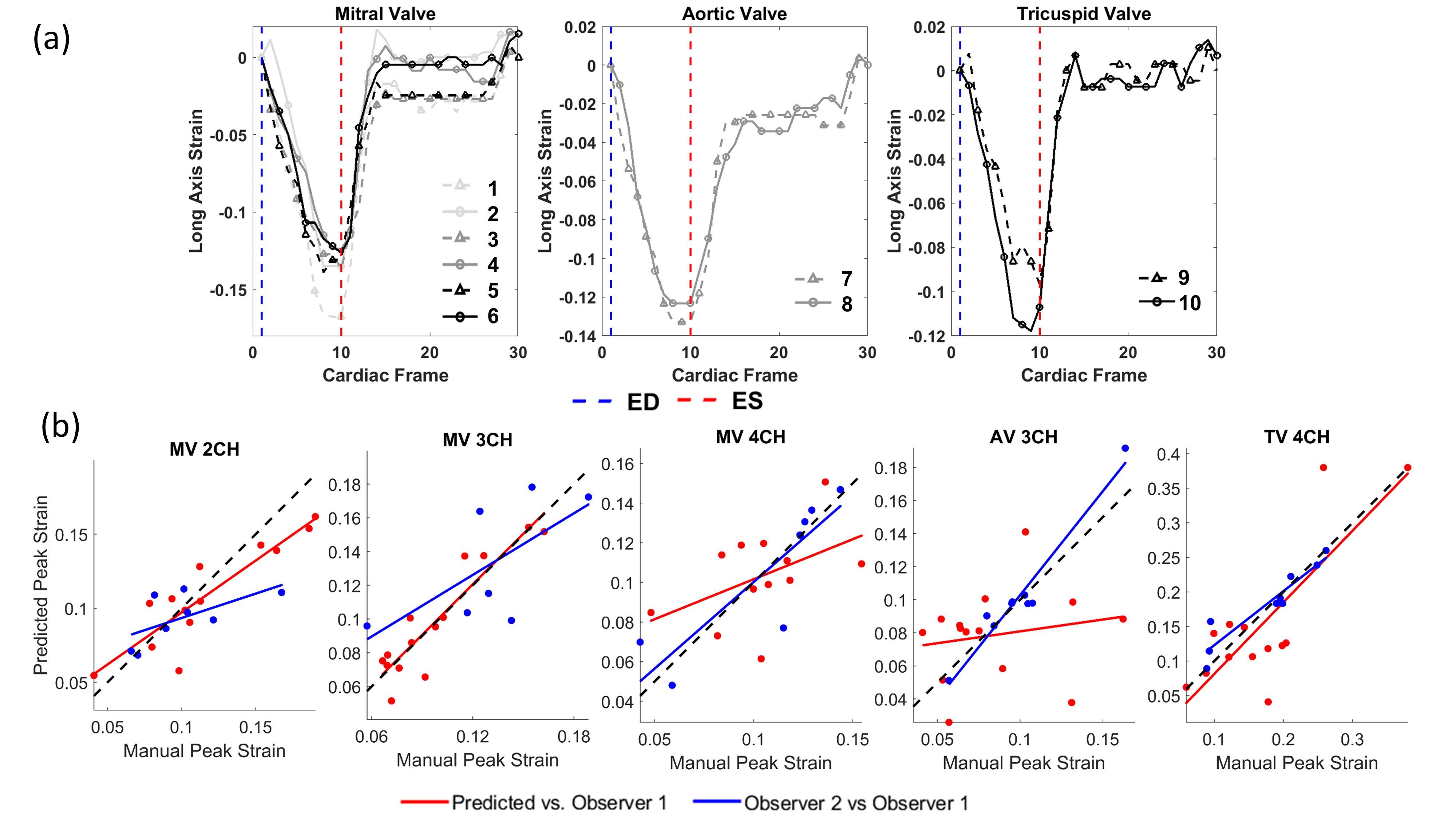}
\caption{a) Long axis strain measured from predicted points for each valve landmark (1-10), grouping together landmarks for the mitral valve (left), aortic valve (centre) and tricuspid valve (right). This is an example case for a set of images previously unseen by the network. b) Peak long axis strain measured from the midpoint of the annulus landmarks plotted for predicted versus manual (red), two different observers (blue) and a perfect correspondence (black dashed line).}
\label{fig:long-axis-example}
\end{figure}

To illustrate the clinical utility of this prediction tool for rapid identification of valve landmarks, an illustrative plot of long-axis strain is shown in Figure \ref{fig:long-axis-example}a. Global longitudinal strain, an accepted measurement for assessing cardiac function, is plotted for each valve landmark. The strain plots clearly show peak strain at end-systole and enable further quantification of clinical measurements such as peak strain and peak strain rates. Peak strain predicted for each valve view was plotted versus the same peak strain measured from manually annotated landmarks in Figure \ref{fig:long-axis-example}b (red). Similarly, peak strain measured from two different observers are plotted for the cases used to measure inter-observer variability (Figure \ref{fig:long-axis-example}b, blue). A perfect correspondence (y = x) is shown as a black dashed line. The 2CH mitral valve, 3CH mitral valve and 4CH tricuspid valve show good agreement for the predicted versus manual peak strain. The predicted versus manual peak strain showed poor agreement for the aortic valve (as illustrated by the discrepancy between red and black dashed lines). This is unsurprising since, at end-systole, rapid ejection causes flow artifacts in the aorta, making the leaflet landmarks hard to detect. In the small sample size used in this plot, the predicted versus manual peak strain also showed poor agreement for the mitral valve 4CH view. However, in general, this shows that, for certain long axis views, the network was capable of extracting not only valve landmarks but also valuable clinical metrics with similar reproducibility as that obtained between multiple observers. 

\begin{figure}[t]
\centering
\includegraphics[width=\textwidth]{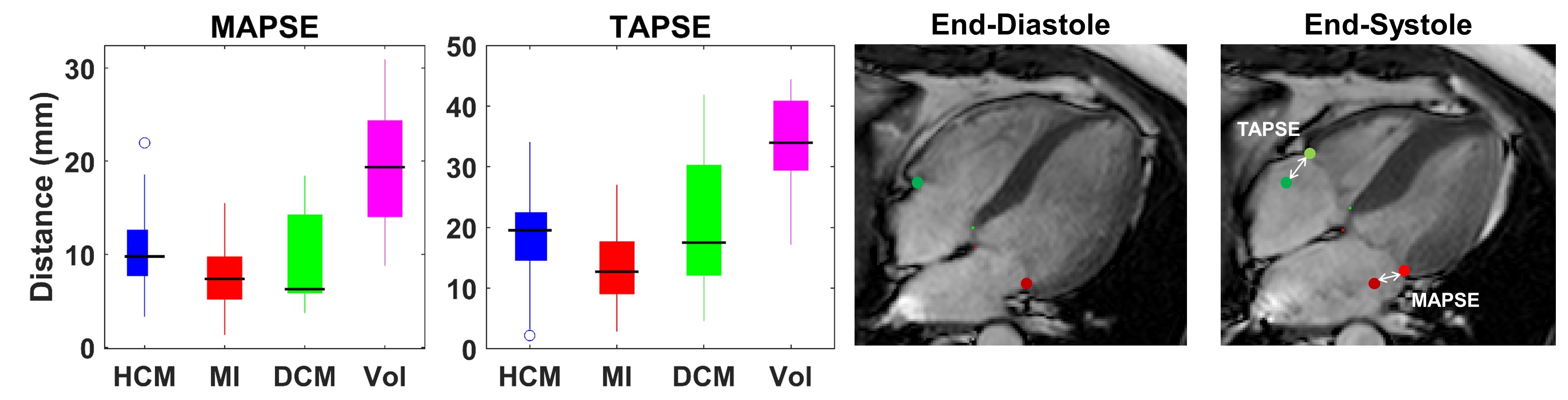}
\caption{Boxplots of MAPSE/TAPSE measured from predicted valve landmarks for each group: HCM, MI, DCM and volunteers.
} 
\label{fig:mapse-tapse}
\end{figure}

MAPSE/TAPSE values (measured from neural network predicted landmarks) plotted in Figure \ref{fig:mapse-tapse} show variations in valve plane motion between patient groups and healthy volunteers. Patients with MI exhibit reduced long-axis strain due to areas of non-contracting myocardium \cite{Hung2010}, which can be seen in these results. MAPSE and TAPSE show large variability within the DCM group, illustrating large variations in phenotype within this patient group. These measurements also reveal that HCM and DCM groups exhibit reduced long-axis motion when compared to healthy volunteers. 

\section{Conclusions}

This study presents a deep learning regression network which utilised standard data augmentation techniques such as flip, rotation, shift and transpose operations in addition to an adaptive minibatch selection process, for the identification of 10 valve landmark points from cardiac long-axis MR images. The network was capable of predicting both the presence and location of these valve landmarks throughout the cardiac cycle in images from a diverse cohort of pathologies. The network robustly predicted valve landmarks with similar or improved accuracy when compared with two easily applied tracking algorithms. 

The network was then applied to predict landmarks in a diverse cohort, including patients with hypertrophic cardiomyopathy, myocardial infarctions, and dilated cardiomypathy, as well as healthy volunteers. These predicted points were used to illustrate differences in clinical measurements such as MAPSE/TAPSE, common measures of systolic function. The landmarks from the network were also used to estimate long-axis strain. Differences between peak long axis strain measured from predicted landmarks versus those measured from manually annotated landmarks were similar to differences seen between multiple observers. 
This tool provides a rapid method for studying, not only valve shape and motion, but also long axis motion in large cohorts with varying pathologies, providing an advance over existing techniques by bypassing the need for manual supervision. 

\subsubsection*{Acknowledgements}
This research was supported by the Wellcome EPSRC Centre for Medical Engineering at the School of Biomedical Engineering and Imaging Sciences, King’s College London (WT 203148/Z/16/Z).

%
%
%
\bibliographystyle{splncs04}

\end{document}